\documentclass[%
aip,
rsi,%
amsmath,amssymb,
reprint,%
]{revtex4-1}

\usepackage{graphicx}% Include figure files
\usepackage{dcolumn}% Align table columns on decimal point
\usepackage{bm}% bold math
\usepackage{algorithm}%
\usepackage{algorithmicx}%
\usepackage{algpseudocode}%
\usepackage{float}
\usepackage[colorlinks=true, linkcolor=blue, citecolor=blue, urlcolor=blue]{hyperref}

\begin{document}
	
	\title{Annealing-inspired training of an optical neural network with ternary weights}
	
	\author{Anas Skalli}
	\email{anas.skalli@femto-st.fr}
	\affiliation{FEMTO-ST Institute/Optics Department, CNRS \& University Franche-Comt\'e, \\15B avenue des Montboucons,
		Besan\c con Cedex, 25030, France}%Lines break automatically or can be forced with \\

	\author{Mirko Goldmann}%
	\affiliation{FEMTO-ST Institute/Optics Department, CNRS \& University Franche-Comt\'e, 
	\\15B avenue des Montboucons,
	Besan\c con Cedex, 25030, France}%

	\author{Nasibeh Haghighi}%
	\affiliation{Technical University of Berlin, Hardenbergstra{\ss}e 36, D-10623 Berlin, Germany}%

	\author{Stephan Reitzenstein}%
	\affiliation{Technical University of Berlin, Hardenbergstra{\ss}e 36, D-10623 Berlin, Germany}%

	\author{James A. Lott}%
	\affiliation{Technical University of Berlin, Hardenbergstra{\ss}e 36, D-10623 Berlin, Germany}%

	\author{Daniel Brunner}
	\affiliation{FEMTO-ST Institute/Optics Department, CNRS \& University Franche-Comt\'e, \\15B avenue des Montboucons,
		Besan\c con Cedex, 25030, France%\\This line break forced% with \\
	}%

	\date{\today}% It is always \today, today,
	%  but any date may be explicitly specified
	
	\begin{abstract}
		
	Artificial neural networks (ANNs) represent a fundamentally connectionnist and distributed approach to computing, and as such they differ from classical computers that utilize the von Neumann architecture. This has revived research interest in  new unconventional hardware to enable more efficient implementations of ANNs rather than emulating them on traditional machines. In order to fully leverage the capabilities of this new generation of ANNs, optimization algorithms that take into account hardware limitations and imperfections are necessary.
	 Photonics represents a particularly promising platform, offering scalability, high speed, energy efficiency, and the capability for parallel information processing. Yet, fully fledged implementations of autonomous optical neural networks (ONNs) with in-situ learning remain scarce. In this work, we propose a ternary weight architecture high-dimensional semiconductor laser-based ONN. We introduce a simple method for achieving ternary weights with Boolean hardware, significantly increasing the ONN's information processing capabilities.
	 Furthermore, we design a novel in-situ optimization algorithm that is compatible with, both, Boolean and ternary weights, and provide a detailed hyperparameter study of said algorithm for two different tasks. Our novel algorithm results in benefits, both in terms of convergence speed and performance. Finally, we experimentally characterize the long-term inference stability of our ONN and find that it is extremely stable with a consistency above 99\% over a period of more than 10 hours, addressing one of the main concerns in the field. Our work is of particular relevance in the context of in-situ learning under restricted hardware resources, especially since minimizing the power consumption of auxiliary hardware is crucial to preserving  efficiency gains achieved by non-von Neumann ANN implementations.

	\end{abstract}
	
	\maketitle
	
\section{Introduction}

Artificial neural networks (ANNs) represent a fundamentally connectionnist and distributed approach to computing, and as such they differ from classical computers that utilize the von Neumann architecture.
 Over the past decade, ANNs have revolutionized computing \cite{lecun2015deep}, disrupting fields ranging from natural language \cite{achiam2023gpt,vaswani2017attention} and image processing \cite{ard2022five} to self-driving vehicles \cite{badue2021self} and game playing \cite{silver2018general}.
 The success of these systems is based on their flexibility, high performance in solving abstract tasks, and their fundamentally parallel approach to information processing, allowing them to distill knowledge from large amounts of data.
 Because they conceptually differ from classical computers, ANNs have to be emulated when running on conventional hardware.
 This has led, on the one hand, to the meteoric rise of more parallel von Neumann processors such as graphical processing units (GPUs) and application specific tensor processing units (TPUs) \cite{reuther2020survey}.
 And, on the other hand, it has spurred significant research interest in developing new hardware to enable more efficient implementations of ANNs \cite{hooker2021hardware}, often leveraging  the strengths of unconventional platforms to either realize co-processors or to build autonomous hardware that directly maps ANN topologies onto the physical substrate.\hfill\break

Although optical neural networks (ONNs) were already demonstrated decades ago \cite{psaltis1988adaptive,psaltis1988multilayered}, photonics has risen again as a particularly promising platform \cite{mcmahon2023physics,abreu2024photonics}, offering scalability \cite{dinc2020optical,rafayelyan2020large,moughames2020three}, high speed \cite{shen2017deep,brunner2013parallel,chen2023deep}, energy efficiency \cite{miller2017attojoule}, and the capability for parallel information processing\cite{brunner2013parallel,lupo2023deep}. Recent advances in photonic hardware include on-chip integrated tensor cores \cite{feldmann2021parallel}, and high dimensional optical pre-processors \cite{wang2023image,xia2023deep}. Among these, semiconductor lasers have emerged as major candidates to implement ONNs due to their ultra-fast modulation rates and complex dynamics\cite{brunner2013parallel}. Vertical-cavity surface-emitting lasers (VCSELs) are notable for their efficiency, speed, intrinsic nonlinearity, and mature complementary metal-oxide-semiconductor (CMOS) compatible fabrication process \cite{muller20111550,vatin2018enhanced}.\hfill\break

Reservoir computing (RC) \cite{jaeger2001echo,maass2002real} simplifies the use of recurrent neural networks (RNNs) by eliminating the need for intensive back-propagation through time training.
In RC, input data is transformed through a high-dimensional network of fixed, interconnected nonlinear nodes—the reservoir, and only the output weights are trained.
Thus, RC and its closely related extreme learning machine \cite{ortin2015unified} can be thought of as the lowest complexity architecture for ANNs, and as such has been implemented on a wide variety of physical substrates \cite{tanaka2019recent} ranging from electronics \cite{appeltant2011information} and spintronics \cite{markovic2019reservoir}, to optics for ultra-fast information processing using a time-multiplexed approach in \cite{brunner2013parallel,appeltant2011information}, frequency-multiplexed approach in \cite{lupo2023deep}, and spatial multiplexing in \cite{skalli2022computational,porte2021complete}, and mechanical substrates \cite{nakajima2013soft}.
Recently, there are also efforts towards a quantum RC implementation to leverage the exponential scaling of the Hilbert space as a high dimensional resource for information processing \cite{markovic2020quantum}.\hfill\break

With the objective of leveraging unconventional ANN hardware to solve a particular task, there has been a significant advancement in hardware-compatible training algorithms \cite{wright2022deep,momeni2024training}. These training methods include backpropagation via a digital twin, i.e., model-based methods, as well as model-free or black-box approaches. Yet, fully realized hardware implementations of these techniques remain rare. For unconventional neural networks to be truly competitive, implementing in-situ training techniques is essential to overcome numerous bottlenecks and to reduce dependence on external high-performance computers, whose usage generally challenges the overall benefit of unconventional ANN computing substrates in general \cite{brunner2021competitive}.\hfill\break

In \cite{porte2021complete,skalli2022computational}, we demonstrated RC through spatial multiplexing of modes on a large area VCSEL (LA-VCSEL), creating a parallel and autonomous network that minimizes the need for an external computer via hardware learning rules with Boolean weights. Here, we crucially expand on our previous work by introducing a minimalist implementation of ternary weights that requires no physical changes to existing hardware and can be applied broadly to other systems relying on Boolean weights.
Boolean weights are appealing as they are exceptionally stable and do not require high-complexity control. Furthermore, using bistable systems such as ferro-electric devices offers a clear avenue towards non-volatile weights, with the associated substantial gains in energy efficiency and system simplicity.
Ternary weights provide a richer representation, which substantially improves the learning capability of the network with minimal demands placed on system complexity and energy cost. 
As a consequence, there has been growing interest in applications of ternary weights in ANNs for increased efficiency, for instance in the context of large language models \cite{zhu2024scalable}. And in the context of electronic hardware implementations \cite{hirtzlin2020digital,laborieux2020low}.
We demonstrate a significant performance increase using ternary weights on th eModified National Institute of Standards and technology database (MNIST) dataset, gaining on average $7\%$ classification accuracy with our fully hardware implemented and parallel optical neural network (ONN) comprising ${\sim}450$ neurons. Additionally, we present an original annealing-like algorithm compatible with both Boolean and ternary weights, which enhances performance and improves convergence speed. Finally, we measure the inference stability of our ONN, demonstrating extremely stable performance over a duration of more than 10 hours, and hence address one of the major concerns of full hardware implementation of semiconductor laser neural networks.

\section{Experimental ONN}

\subsection{Optical experimental setup}
\begin{figure}[h!]
\begin{center}
\includegraphics[width=1\linewidth]{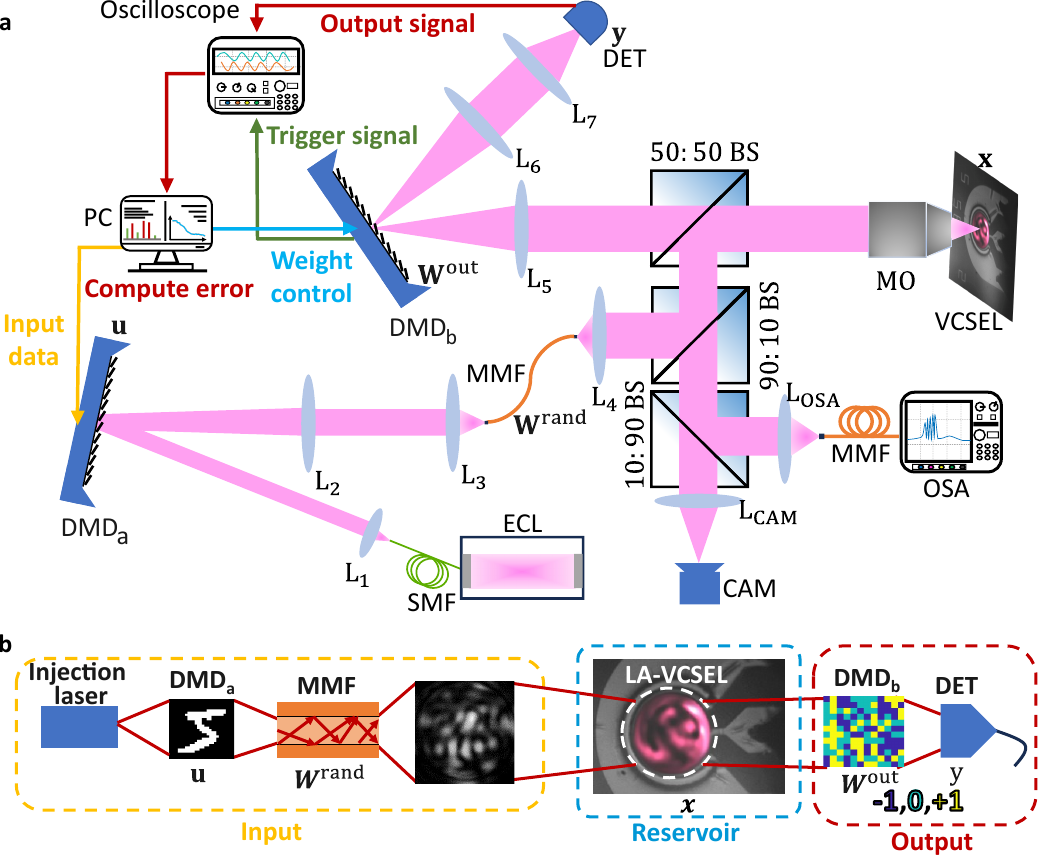}  
\caption{\textbf{Experimental setup and concept scheme.} \textbf{a} Schematic of the full experimental setup. Details about the individual components are given in table \ref{tab:components_ref}. \textbf{b} Simplified diagram explaining our experimental scheme for clarity.}
\label{fig:schematic_scheme}
\end{center}
\end{figure}

Our ONN experiment builds upon our previous implementations based on LA-VCSELs presented in \cite{porte2021complete,skalli2022computational}. As such, it follows the RC architecture, and is therefore comprised of three parts: the input layer, the reservoir and the output layer. The complete experimental setup schematic is shown in Fig. \ref{fig:schematic_scheme}(a), and a simplified diagram structure, shown for comparison in Fig. \ref{fig:schematic_scheme}(b), illustrates the conceptual analogy between machine learning concept and optical hardware. Table \ref{tab:components_ref} in appendix \ref{secA1} gives a detailed list of all the components used.\hfill\break

The input layer consists  of three main components, a continuously tunable external cavity laser (ECL, Toptica CTL 950) that we use to carry information, a digital micro-mirror device (DMD, Vialux XGA 0.7" V4100), and a multimode fibre (MMF,  THORLABS M42L01). First, the single-mode fiber-coupled ECL is collimated with an $f_1=6\ \text{mm}$ focal distance lens ($L_1$, THORLABS C110TMD-B), yielding a collimated beam diameter of ${\sim} 1.4\ \text{mm}$, that is used to illuminate $\text{DMD}_\text{a}$, which through its pixels modulates the input beam that carries information to the LA-VCSEL. A DMD is a digitally controlled matrix of micro-mirrors, which can flip between two angles, here at angles of $\pm 12^\circ $, allowing us to display Boolean images that constitute our input information $\mathbf{u}$. We then use a simple two-lens imaging system $L_{2}$ (THORLABS AC508-150-B-ML, $f_{2} = 150~ \text{mm}$), and $L_{3}$ (THORLABS C110TMD-B, $f_{3} = 6~ \text{mm}$) adjusting the MMF's image size to approximately the size of the collimated ECL beam on $\text{DMD}_\text{a}$. This ensures maximal coupling into the $50~\mu\text{m}$ diameter MMF, which through its transmission matrix passively implements the random input weights $\mathbf{W}^{\text{rand}}$ for RC.\hfill\break

Following this, the nearfield output of the MMF, $\mathbf{W}^\text{rand}\mathbf{u}$, is imaged onto the LA-VCSEL. Here we use a LA-VCSEL with an aperture of ${\sim}55~\mu\text{m}$ and a threshold current of $I_{\text{th}}=20~\text{mA}$. A standard ground-signal-ground (GSG) probe (Microworld MWRF-40A-GSG-200-LP) is used to bias the LA-VCSEL, and the laser is set to ${\sim} 28^\circ~\text{C}$ with a thermal stability on the order of $10~\text{mK}$. The imaging system we use here is also a two lens system. The output of the multimode fibre is collimated using $L_4$  (THORLABS AC127-20-B-ML, $f_{4} = 20~ \text{mm}$) and imaged onto the large area VCSEL using a microscope objective (MO, OLYMPUS LMPLN10XIR, $f_{\text{MO}} = 18~\text{mm}$). We achieve a magnification factor of $m = f_{\text{MO}} / f_{4} = 0.9$ , closely matching the MMF's and the LA-VCSEL's apertures. Furthermore, the MMF's numerical aperture (NA) should be chosen such that, considering the effects of magnification $m$, it it  is similar but smaller than that of the LA-VCSEL. 
In our design, the LA-VCSEL's physical properties and dynamics form the reservoir's components, including nonlinear nodes and their interconnections. These nodes, represented by specific areas on the LA-VCSEL's surface, interact through inherent physical mechanisms within the device. Localized Gaussian-like coupling arises from carrier diffusion in the semiconductor quantum wells, while a more complex global coupling is created by the optical field's diffraction within the laser cavity. 
The LA-VCSEL takes the input information $\mathbf{W}^\text{rand}\mathbf{u}$ and transforms it in a complex nonlinear way as is generally the case for optical injection into a semiconductor laser \cite{hicke2013information}. This process produces the reservoir state $\mathbf{x}$, and the complexity of the process can be appreciated from the perturbed LA-VCSEL mode profile under injection shown in  Fig. \ref{fig:schematic_scheme}(b).\hfill\break

Finally, the output layer is realized via $\text{DMD}_\text{b}$ and a photodetector (DET, Thorlabs PM100A, S150C). By imaging the LA-VCSEL's near field onto $\text{DMD}_\text{b}$, we discretize the VCSEL's continuous spatial nonlinear response into a discrete matrix, sampling its surface and thus allowing us to adjust the effective neuron-count for one and the same device through super-pixel size control.\par

Imaging of the LA-VCSEL onto $\text{DMD}_\text{b}$ is realized via a two-lens system, using $L_{5}$ (THORLABS AC254-100-B-ML, $f_{5} = 100~\text{mm}$) and MO. The choice of $L_{5}$ is motivated by the size of $\text{DMD}_\text{b}$'s pixels and the characteristic size of features in the LA-VCSEL's mode profile. Ideally, we want to oversample the mode profile with the DMD pixels, which we ensure via magnification $m = f_{5} / f_{\text{MO}} = 5.55$. The size of speckles on the LA-VCSEL's surface are ${\sim}5.6~\mu \text{m}$, while the size of one mirror on $\text{DMD}_\text{b}$ is ${\sim}13.7~\mu \text{m}$, and taking into account the magnification every speckle is imaged onto $5$ mirrors.
The LA-VCSEL is magnified on $\text{DMD}_\text{b}$ resulting in a diameter of $305~\mu\text{m}$ corresponding to an area of $24\times24$ DMD mirrors. $L_{6}$ (THORLABS AC254-150-B-ML, $f_{6} = 150~\text{mm}$) and $L_{7}$ (THORLABS AC254-45-B-ML, $f_{7} = 45~\text{mm}$) are chosen to demagnify the LA-VCSEL's image to ensure it fits on the DET.\par

The total area of the rectangular region of interest on $\text{DMD}_\text{b}$ represents $24 \times 24 = 576$ mirrors in total. Since the LA-VCSEL is circular, its area corresponds to $576 \cdot \pi / 4 \simeq 452 $ mirrors. We hence achieve approximately 450 fully parallel neurons. The DMD's pixels act as a spatial filter between the LA-VCSEL and the DET, realizing output weights
$\mathbf{W}^{\text{out}}$ by directing the optical signal associated to $\text{DMD}_\text{b}$ mirrors to one of the two angular configurations one of which has DET, while signals associated to $\text{DMD}_\text{b}$ mirrors in the opposite angular configuration are discarded. $\text{DMD}_\text{b}$ therefore implements a Boolean readout matrix. Through $\text{DMD}_\text{b}$, we then tune the spatial positions of the LA-VCSEL that contribute to the optical power detected at DET, and by optimizing this mirror configuration, we train the output $\mathbf{y}$ of the reservoir. Although the DMD is an inherently Boolean, i.e. binary device, we will later in section \ref{sec:trinary_implem} explain how we use it to implement ternary weights.\hfill\break

As shown in Fig. \ref{fig:schematic_scheme}(a), a digital computer controls our experiment, sending the input data to be written on $\text{DMD}_\text{a}$, recording the output via an oscilloscope as well as controlling the output weights on $\text{DMD}_\text{b}$. This computer only acts as a supervisor to our experiment and to implement learning, and as such does not partake in any information processing. It could therefore be replaced by a low-power alternative such as a single board compunter like the Raspberry Pi.\hfill\break

The data acquisition loop or forward pass of our ONN is as follows. A batch of $N$ images is loaded on $\text{DMD}_\text{a}$'s onboard memory to allow for a fast frame rate of $15~\text{kHz}$. The computer then triggers $\text{DMD}_\text{a}$, which in turn hardware triggers the oscilloscope to start the acquisition of the signal detected at DET. Each frame on $\text{DMD}_\text{a}$ is displayed for $66~\mu \text{s}$, which is orders of magnitude slower than the intrinsic time scales of the LA-VCSEL (${\sim}1~\text{ns}$), meaning that we operate the LA-VCSEL in its steady state. The waveform acquired via the oscilloscope is digitally downsampled  to $N$ points, yielding the output of the ONN $\mathbf{y}^{\text{out}}$, allowing us to compute an error between this output and a set target $\mathbf{y}^{\text{target}}$, according to the normalized mean square error (NMSE) computed at each epoch $k$:

 \begin{equation}
     \operatorname{NMSE}_{k} =\frac{1}{N \times \operatorname{std}(\mathbf{y}_{k}^{\text{out}})}\sum_{i=1}^{N}(\mathbf{y}_{k}^{\text{out}}(i) - \mathbf{y}^{\text{target}}(i))^2.
 \end{equation}

\begin{figure}[h!]
\begin{center}
\includegraphics[width=1\linewidth]{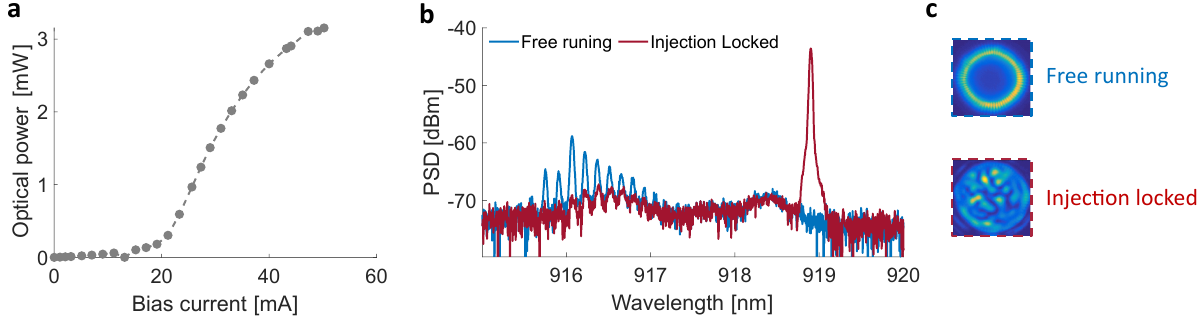}  
\caption{\textbf{Experimental charaterization of the LA-VCSEL.} \textbf{a} L-I curve of the LA-VCSEL. \textbf{b} Spectra of the free running (blue) and injection locked (red) LA-VCSEL. \textbf{c} Mode profiles of the free running (blue) and injection locked (red) LA-VCSEL.}
\label{fig:figure_charac}
\end{center}
\end{figure}

Following the detailed characterization provided in \cite{skalli2022computational} and to ensure optimal performance, we operate the LA-VCSEL at $I_{\text{bias}} = 1.5 I_{\text{threshold}} \sim 30~\text{mA}$, see Fig. \ref{fig:figure_charac}(a). Moreover, the injection wavelength $\lambda_{\text{inj}} \sim 919~\text{nm} $ was optimized to yield optimal injection locking conditions as shown in Fig. \ref{fig:figure_charac}(b), resulting in maximal suppression of the LA-VCSEL's numerous modes when left free, and the injected power is set to match the emission power of the LA-VCSEL yielding an injection power ratio $\operatorname{PR} = P_{\text{inj}} / P_{\text{VCSEL}} \sim 1$. The free-running and injection locked mode profiles of the LA-VCSEL are shown in  Fig. \ref{fig:figure_charac}(c).

\subsection{Ternary weights implementation\label{sec:trinary_implem}}

A simple, yet powerful change to our previous experimental setup was the implementation of ternary weights, i.e $ W^{\text{out}} \in \{-1, 0, +1\}$
with virtually no change to the experiment. 
Let \( \mathbf{W}^{\text{out}} \) be our output matrix with ternary entries. We can then define two strictly positive matrices, \( \mathbf{W}^{\text{out+}} \) and \( \mathbf{W}^{\text{out-}} \) as follows:

$$
\mathbf{W}_{ij}^{\text{out+}} = \begin{cases} 
1 & \text{if } \mathbf{W}_{ij}^{\text{out}} = 1, \\
0 & \text{otherwise}.
\end{cases}
$$

$$
\mathbf{W}_{ij}^{\text{out-}} = \begin{cases} 
1 & \text{if } \mathbf{W}_{ij}^{\text{out}} = -1, \\
0 & \text{otherwise}.
\end{cases}
$$

\begin{figure}[h!]
\begin{center}
\includegraphics[width=0.75\linewidth]{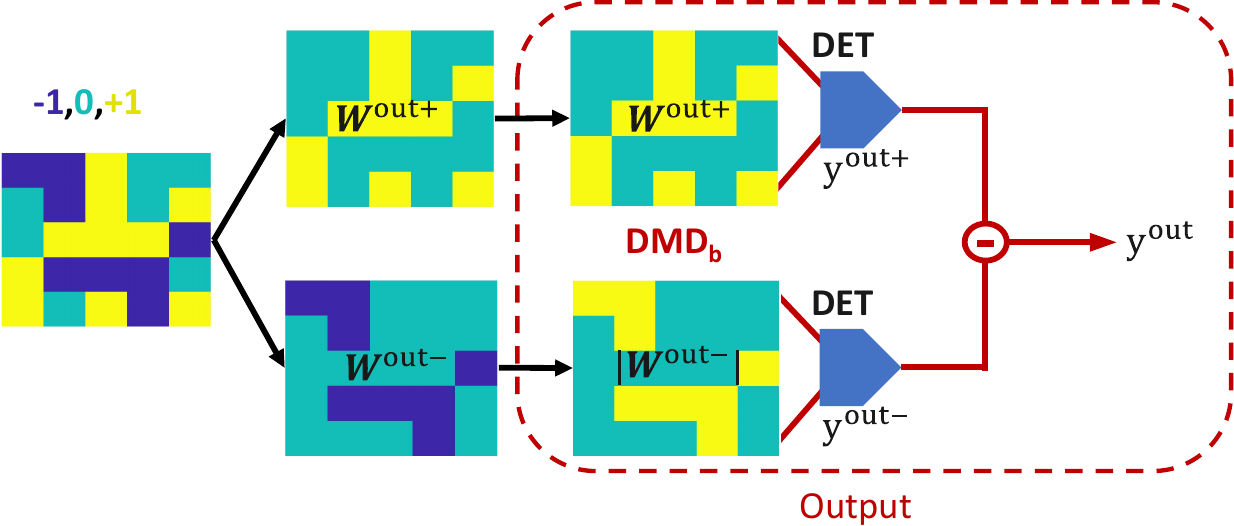}  
\caption{\textbf{Ternary weights implementation.} Diagram showing how we use a single DMD to achieve ternary weights.}
\label{fig:trinary_diagram}
\end{center}
\end{figure}

By sequentially displaying \( \mathbf{W}^{\text{out+}} \) and \( \mathbf{W}^{\text{out-}} \) and measuring their respective outputs signals, $\mathbf{y}^{\text{out+}}$ and $\mathbf{y}^{\text{out-}}$, the output of our ONN, $\mathbf{y}^{\text{out}}$, is computed by electronic subtraction $\mathbf{y}^{\text{out}} = \mathbf{y}^{\text{out+}} - \mathbf{y}^{\text{out-}}$. Note that, both, \( \mathbf{W}^{\text{out+}} \) and \( \mathbf{W}^{\text{out-}} \), are positive Boolean matrices that by definition can be displayed on the DMD. The negative weights result from the subtraction of their respective output signals. Figure \ref{fig:trinary_diagram} gives a diagram view of these operations.

%Training is realized via the same simulated annealing-like algorithm as presented in Chapter 4, with the only difference being that the weights are now trinary. The implication being that at each epoch instead of flipping the state of previously Boolean weights, it is assigned a value of either ${-1,0, \text{or} +1}$ with equal probability.

However, in its present form, our implementation of ternary output weights results in halving the inference bandwidth due to its sequential nature, requiring two measurements at each step. 
To remedy this, we could create an optical copy of the LA-VCSEL state on top of $\text{DMD}_\text{b}$. Each state would be sent to a separate area of $\text{DMD}_\text{b}$, where the corresponding weight matrices \( \mathbf{W}^{\text{out+}} \) and \( \mathbf{W}^{\text{out-}} \) would be displayed simultaneously and two outputs would be detected by separate detectors, and their outputs would be subtracted electronically in real-time. This would allow us to maintain the same inference bandwidth as with Boolean weights, while still benefiting from the performance increase provided by ternary weights. A similar concept was already leveraged in the first experimental ONN demonstration \cite{farhat1985optical}.

%\section{Improved hardware learning (alpha strategy) \label{sec:learning}}
\subsection{In-situ learning\label{sec:learning}}

Our in-situ optimization algorithm builds upon, and significantly improves the evolutionary algorithm presented in \cite{bueno2018reinforcement,andreoli2020boolean} designed for Boolean learning. Originally, a single randomly chosen Boolean weight is inverted at each epoch, and if this change results in a decreased error it is kept, otherwise we revert the Boolean weight to its former value then select a different Boolean weight.
Here, instead of simply inverting a single Boolean weight, i.e flipping a mirror, we propose flipping several mirrors per epoch. Furthermore, we make the number of mirrors we flip $n^{\text{mirrors}}$ adaptive and proportional to the error according to 
\begin{equation}
    n^{\text{mirrors}} = \lceil \alpha \cdot \operatorname{NMSE}_{k} \rceil,
\end{equation}
where $\alpha$ is a hyperparameter that can be likened to a learning, that does not control step size but rather the number of permutations from one epoch to the next. Setting $\alpha = 0$ corresponds to the original algorithm, the ceiling function $\lceil \ \rceil$ ensures that at least one mirror is flipped at each epoch. The pseudo-code for our algorithm and optimization loop is given in algorithm \ref{alg:alpha_strat}.\par

When using ternary weights, at each epoch instead of simply flipping the state of previously Boolean weights, they are individually assigned a value of either ${-1,0, \text{or} +1}$ with equal probability, making our simple algorithm inherently compatible with Boolean and ternary weights.\par

This strategy combines elements of random search and a well known gradient free optimization technique known as simulated annealing \cite{from1987simulated,kirkpatrick1983optimization,bertsimas1993simulated}. Indeed, simulated annealing iteratively optimises a function by applying random fluctuations to the given parameters while only accepting changes that yield positive outcomes in a process similar to a random walk. Crucially, the magnitude of these random fluctuations depends on a so-called temperature which decreases with time, causing the algorithm to settle at a given position in the landscape. Therefore, our strategy can be viewed as a version of simulated annealing where instead of arbitrarily relying on time to force convergence we directly leverage the error to guide our search in the landscape. \par

\begin{algorithm}[H]

\caption{Ternary Weights adaptive optimization\label{alg:alpha_strat}}
\begin{algorithmic}[1]
\State Initialize $W^{\text{out}}$ randomly
\State $W^{\text{out\_best}} \gets W^{\text{out}}$
\State $\text{MSE\_best} \gets \text{Forward\_pass}(W^{\text{out}})$

\While{not converged}
    \State $n^{\text{mirrors}} \gets \lceil \alpha \times \text{MSE\_best} \rceil$
    \State $W^{\text{out\_temp}} \gets W^{\text{out}}$
    \For{$i = 1$ to $n^{\text{mirrors}}$}
        \State Randomly select a mirror $m_i$
        \State Randomly set $W^{\text{out\_temp}}[m_i] \in \{-1, 0, +1\}$
    \EndFor
    \State $\text{MSE\_temp} \gets  \text{Forward\_pass}(W^{\text{out\_temp}})$

    \If{$\text{MSE\_temp} < \text{MSE\_best}$}
        \State $W^{\text{out}} \gets W^{\text{out\_temp}}$
        \State $W^{\text{out\_best}} \gets W^{\text{out\_temp}}$
        \State $\text{MSE\_best} \gets \text{MSE\_temp}$
    \Else
        \State Revert to $W^{\text{out}}$
    \EndIf
\EndWhile
\State \Return $W^{\text{out\_best}}$

\end{algorithmic}
\end{algorithm}

\section{Results}
\subsection{Benchmark tasks}

In order to benchmark our ONN we use two tasks. The first is binary header recognition, the second the is hand-written MNIST dataset. Due to the scalar output given by the single photodetector DET, we use "one-vs-all" classification. For our tests, input sequences $\mathbf{u}$ where comprised of $N=1000$ images comprised of a 50-50 split between positive and negative examples for one-vs-all classification.

As a reference, for the first task the input data consists of binary pie-shaped headers. The Gaussian input beam is divided into $n_{\text{bits}}$ 
equal sections that encode the header. While seemingly simple, these orthogonal patterns are quite convenient and allow us to reliably scale the dimensionality and complexity of our dataset and computational task. A more detailed description of the encoding used for the binary header can be found in \cite{skalli2022computational,porte2021complete}.

\subsection{Influence of alpha}

\begin{figure}[h!]
\begin{center}
\includegraphics[width=1\linewidth]{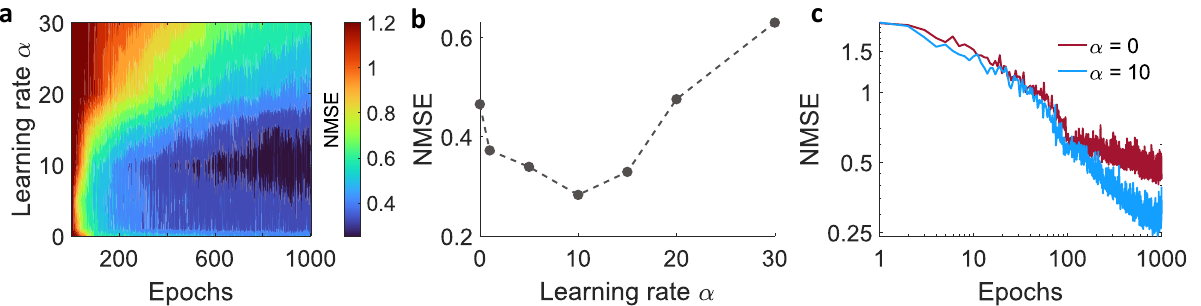}
\caption{\textbf{Impact of hyperparameter $\alpha$ on the learning performance for a 4-bit header recognition task.}}
\label{fig:alphascan_4bit}
\end{center}
\end{figure}

First let us study the influence of $\alpha$ on convergence. Figure \ref{fig:alphascan_4bit} shows the impact of $\alpha$ on the performance and convergence for a 4-bit header recognition task. The error decreases faster for an optimal value of $\alpha=10$, yet higher values of $\alpha$ lead to a more unstable learning trajectory. Indeed, flipping more mirrors is akin to bigger steps taken in the search space at each epoch, as such, $\alpha$ behaves much like a learning rate in the context of gradient descent. Moreover, the errors reached by optimal values of $\alpha $ are lower. With $\alpha = 10$ we have $\operatorname{NMSE} = 0.25$  compared to $0.5$ for the original algorithm with $\alpha = 0 $. This shows that the adaptive strategy is a simple yet powerful modification to the original algorithm. We should note that $\alpha$ is task dependant. Leveraging this improved strategy we are able to increase performance and reach a symbol error rate (SER) of $1.5\%$ for a 6-bit header recognition task.

Rather than using the synthetic benchmark on binary header recognition, the MNIST task provides a more challenging and genuine image classification problem. Here we study the influence of $\alpha$ on convergence using one-vs-all classification of digits 8 and 9, the hardest in the dataset. Generally speaking, the same trend appears with optimal $\alpha$ values in the range $5\leq \alpha \leq 20$. Choosing optimal $\alpha$ values results in faster convergence and better performance, allowing the ONN to reach ${\sim} 85 \%$ accuracy for $\alpha=20$  instead of ${\sim} 81 \%$ for $\alpha=2$ while converging in ${\sim} 200$ epochs as opposed to ${\sim} 600$.\hfill\break

To conclude, our adaptive search strategy offers significant gains both in terms of performance and convergence speed with minimal computational overhead with no changes to the existing hardware, while being compatible with both Boolean and ternary weights.

\begin{figure}[h!]
\begin{center}
\includegraphics[width=1\linewidth]{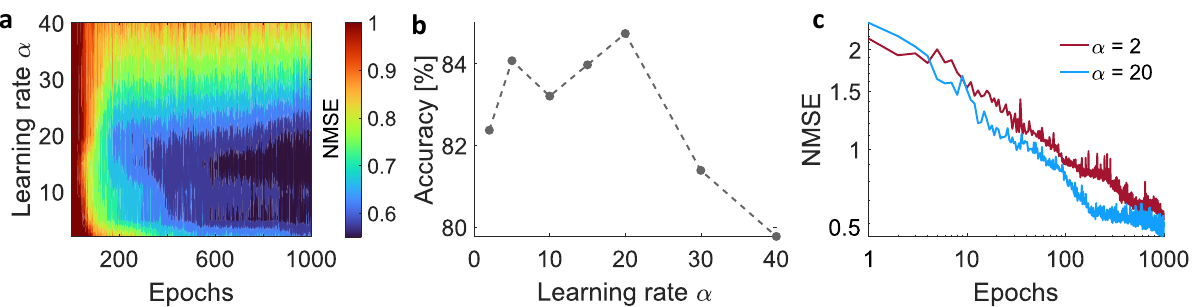}
\caption{ \textbf{Impact of tuning hyperparameter $\alpha$ on the learning performance for classification of digit 8 in the MNIST dataset.}}
\label{fig:alphascan_MNIST}
\end{center}
\end{figure}

\subsection{Benefits of ternary weights  \label{sec:trinary}}

\begin{figure}[h!]
\begin{center}
\includegraphics[width=1\linewidth]{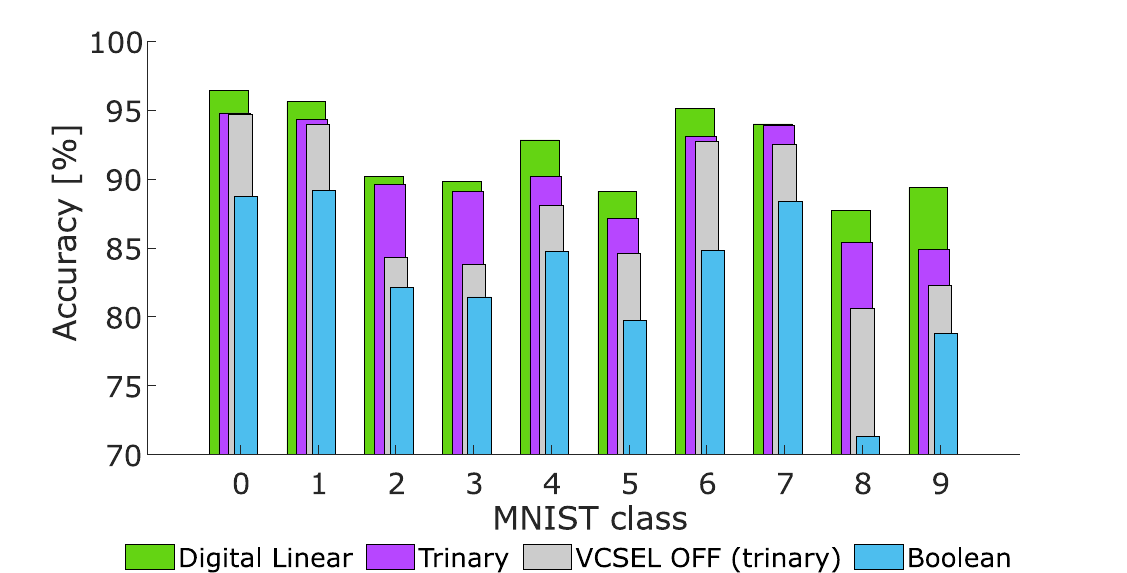}
\caption{\textbf{Performance on the MNIST dataset for Boolean and ternary weights compared to a digital linear classifier.}}
\label{fig:bool_vs_3val}
\end{center}
\end{figure}

We then experimentally quantify the net gain following the implementation of ternary weights. Figure \ref{fig:bool_vs_3val} shows the performance for one-vs-all classification via our ONN for every digit in the MNIST dataset for Boolean and ternary weights. 
%As a reminder, in one-vs-all classification, for a task containing $n$ classes then we train $n$ independent binary classifiers. 
In addition, it also shows the performance for a digital linear classifier trained with ridge regression on the same data. 
As a reminder we used training and testing sequences $\mathbf{u}$ of length $N=1000$ images. Crucially, Boolean weights perform poorly, achieving on average ${\sim}83.5\%$ classification accuracy on the testing dataset. 
In contrast, with no physical changes to our experimental setup, we can achieve $90.4\%$ on average with ternary weights approaching the digital linear limit of ${\sim} 91.8\%$. 
The performance of our ONN with the LA-VCSEL switched off was also measured with ternary weights as a reference and reaches ${\sim} 87.2\%$. This corresponds to a  more linear hardware system comprised only of the MMF as a passive linear mixing element and the absolute square nonlinearity of optical detection. 
%In principle, even with the LA-VCSEL switched off, there still is a nonlinearity present in the physical system. 
%Indeed, our output detector by measuring the light's intensity $I^{\text{out}} = \sum_{i} W_i^{\text{out}}*|E_i^{\text{VCSEL}}|^2$ implements a nonlinear operation on the MMF field. 
Yet, as shown in our present measurement, this nonlinearity is not sufficient and cannot result in performance close to a digital linear classifier. 
Interestingly, our ternary weight ONN with the LA-VCSEL switched off performs better than when the LA-VCSEL is on and using Boolean weights, which means that at this very low-end, weight resolution is a severely limiting factor. Conceptually, because they are positive, Boolean weights cannot exploit nodes or neurons whose responses are anti-correlated with the desired target. The respective weights for these nodes are set to 0 when using Boolean weights which effectively restricts the dimensionality of our ONN.\hfill\break

This comparison highlights how crucial weight resolution and sign can be. Although the performance improvements when using ternary weights are substantial, our ONN still falls short and cannot outperform a digital linear classifier. 
%It therefore follows that our next step should be to further improve our experimental setup, in order to overcome the digital linear limit. A rather non-invasive way of improving our current experiment would be to increase the imaging magnification of the LA-VCSEL on $\text{DMD}_\text{b}$ by changing $L_{5}$. Doing so would effectively provide higher resolution weights since every LA-VCSEL speckle would be significantly oversampled. Another way would be to replace $\text{DMD}_\text{b}$ with a spatial light modulator (SLM), the vast majority of which provide at least 8-bit resolution without the need of additional magnification. Moreover, through the use of phase modulation in a 4-f configuration and spatial filtering at the detection, one can achieve positive and negative weights in a single measurement.

\subsection{Long-term stability characterization}

The ternary-weight ONN described above performs all processing steps online, reducing the need for additional offline resources. In addition, due to its full hardware implementation, our system is prone to drifts.  This makes long-term stability and robustness against drift crucial in our setup. 
Importantly, these drifts are significantly mitigated through standard proportional-integral-derivative (PID) temperature control of the LA-VCSEL and mechanically securing the injection MMF in place. 
In our long-term characterization, we find that post learning convergence, the system maintains stable performance for several hours, exhibiting only gradual performance degradation rather than abrupt drops. This stability suggests that continuous, online learning can effectively counteract these slow drifts, maintaining consistent system performance over time.\par

\begin{figure}[h!]
\begin{center}
\includegraphics[width=1\linewidth]{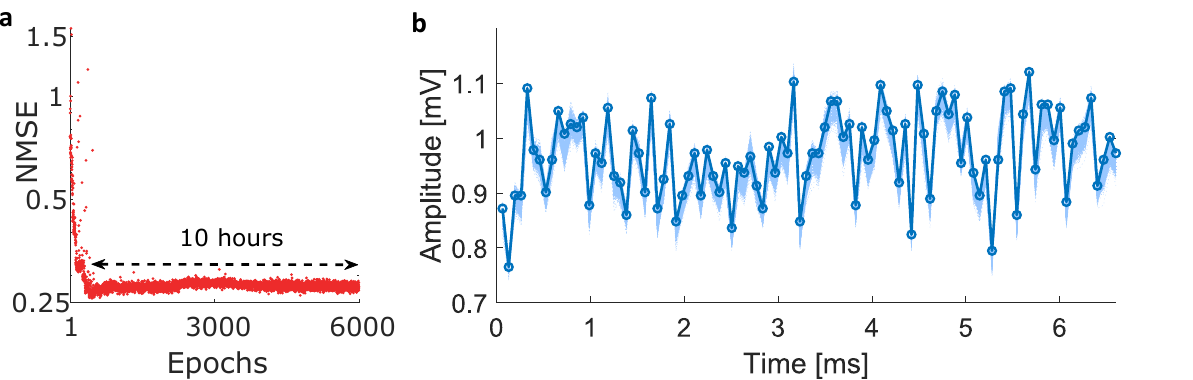}
\caption{\textbf{Long-term stability characterization after learning.} \textbf{a} NMSE learning curve, after convergence, output weights are fixed and the output is mesured every 10 seconds over a period of 10 hours. \textbf{b} Output time traces over 10 hours, the first output is shown in bold while following outputs are in a lighter shade. The correlation between responses is ${\sim}99.3\%$.}
 
\label{fig:long_term_drift}
\end{center}
\end{figure}

In order to quantify the impact of physical drift on our PNN's performance, we first conducted an optimization loop for 100 epochs, training the ONN to classify digit 0 of MNIST. After convergence to a low error, we kept the output weights fixed and countinuously monitored the output of the system every 10 seconds over a period of 10 hours. We were therefore able to measure long-term physical drifts and their impact on computing performance in our system, see Figure \ref{fig:long_term_drift}(a). Surprisingly, the error remains effectively constant over 10 hours with no noticeable drift in terms of performance. Finally, Fig. \ref{fig:long_term_drift}(b) shows the raw output of our setup during these 10 hours, the dark blue line shows the first response at the beginning of the 10 hours, and the light blue shows the next responses over the duration of the experiment. Importantly, the correlation between responses, i.e. consistency \cite{kanno2012consistency}, after learning is ${\sim}99.3\%$, showcasing how even with our simple control schemes we can greatly minimize instabilities in our system.

It should be noted that, because the inference speed is orders of magnitude slower than the dynamics of the VCSEL, we do not measure the dynamical stability of our device on the timescale of its inherent dynamics. We rather measure a time averaged or steady state response output of the LA-VCSEL, which toghether with injection locking further increases stability.

\clearpage
\section{Conclusion}

We significantly improved upon our previous LA-VCSEL based ONN, by demonstrating an efficient and low complexity method to implement ternary weights that is broadly compatible with Boolean-weight based RC. Crucially, we report significant improvements in performance with no physical changes to our experimental setup.\par
In addition, we introduced a novel in-situ optimization algorithm that is compatible with, both, Boolean and ternary weights. We provided a detailed hyperparameter study of said algorithm for two different tasks, and experimentally verified its benefits in terms of convergence speed and performance gain, resulting in an increase of $7\%$ on average when going from Boolean to ternary weights. We also confirmed that weight resolution is the main limiting factor in our relatively low neuron-count ONN, offering a clear avenue for future improvements. Finally, we experimentally characterized the long-term inference stability of our ONN and found that it was extremely stable with a consistency above $99\%$ over a period of more than 10 hours. Our work is of high relevance in the context of in-situ learning particularly under restricted hardware resources.

\section*{Declarations}

The authors acknowledge the support of the Region Bourgogne Franche-Comt\'{e}. This work was supported by the EUR EIPHI program (Contract No. ANR-17-EURE-0002), by the Volkswagen Foundation (NeuroQNet I\&II), by the French Investissements d’Avenir program, project ISITE-BFC (contract ANR-15-IDEX-03), and by the German Research Foundation (via SFB 787), and by the European Union’s Horizon research and innovation program under the Marie Skłodowska-Curie grant agreement No 860830 (POST DIGITAL) and 101044777 (INSPIRE).

\section*{Author contribution}
AS and DB conceived the experiment and in combination with MG developed the learning algorithm.
NH fabricated and characterized the LA-VCSEL and many neighboring devices under the supervision of JL to select a suitable VCSEL for the experiment using VCSEL wafers designed and provided by JL. SR provided the fabrication infrastructures.
AS carried out all experiments under the supervision of DB,  AS wrote the manuscript with contributions of all authors.

\section{List of optical Components}\label{secA1}

\begin{table}[h!]
    \centering
    \begin{tabular}{c|c}
       Components  &  Reference\\ 
         ECL & TOPTICA CTL 950 \\
         SMF & \\
         $L_{1}$ & THORLABS C110TMD-B)\\
         $DMD_{\text{a}}$ & Vialux XGA 0.7" V4100 \\
         $L_{2}$ & THORLABS C110TMD-B \\
         $L_{3}$ & THORLABS AC508-150-B-ML \\
         MMF & THORLABS M42L01\\
         $L_{4}$ & THORLABS AC127-20-B-ML \\
         BS $90:10$ & THORLABS BSX11R \\
         BS $50:50$ & THORLABS CCM1-BS014/M \\
         MO & OLYMPUS LMPLN10XIR \\
         VCSEL & Custom made in unversity cleanroom\\
         $L_{5}$ & THORLABS AC254-100-B-ML \\
         $DMD_{\text{b}}$ & Vialux XGA 0.7" V4100 \\
         $L_{6}$ & THORLABS AC254-150-B-ML\\
         $L_{7}$ & THORLABS AC254-45-B-ML\\
         DET & THORLABS PM100A, S150C \\
         Oscilloscope & ROHDE SCHWARZ RTO 1084\\
         PC & Standard Dell Precision i5-6400 8Gb RAM\\
         BS $ 10:90$ & THORLABS BSN11R \\
         $L_{\text{OSA}}$ & THORLABS AC254-35-B-ML \\
         MMF & THORLABS M42L01\\
         OSA & Yokogawa AQ6370D\\
         $L_{\text{CAM}}$ & THORLABS AC254-100-B-ML \\
         CAM & IDS-UI-LI348x-E\\
    \end{tabular}
    \caption{Table showing all components in the optical setup for reference.}
    \label{tab:components_ref}
\end{table}

\section{Bibliography}

\bibliography{sn-bibliography}% common bib file
%% if required, the content of .bbl file can be included here once bbl is generated
%%\input sn-article.bbl
%\bibliographystyle{ieeetr}
	
\end{document}